\documentclass[11pt]{article}
\pdfoutput=1

\usepackage[english]{babel}
\usepackage{amsmath}
\usepackage{amssymb}
\usepackage{amsfonts}
\usepackage[ruled,vlined]{algorithm2e}
\usepackage{mathtools, booktabs}
\usepackage{graphicx}
\usepackage[round]{natbib}
\usepackage{appendix, bbm, color}
\usepackage{fullpage, multirow}
\usepackage{authblk}
\usepackage[pdftex]{hyperref}
\usepackage{booktabs}

\usepackage{bbm}
\usepackage{cancel} 
\usepackage{comment}

\usepackage[pdftex,dvipsnames]{xcolor}

\providecommand{\keywords}[1]{\textbf{\textit{Keywords:}} #1}
\begin{document}
\title{Bayesian GARCH Modeling of Functional Sports Data}
\author[1]{Patric Dolmeta}
\author[2]{Raffaele Argiento}
\author[3]{Silvia Montagna\thanks{silvia.montagna@unito.it}}

\affil[1]{Department of Decision Sciences, Universit\`{a} Bocconi, Milano, Italy}
\affil[2]{Department of Statistical Sciences, Universit\`a Cattolica del Sacro Cuore, Milano, Italy \& Collegio Carlo Alberto, Torino, Italy}
\affil[3]{ESOMAS Department, University di Torino, Torino, Italy \& Collegio Carlo Alberto, Torino, Italy}

\maketitle

\begin{abstract}
	
	The use of statistical methods in sport analytics has gained a rapidly growing interest over the last decade, and nowadays is common practice. In particular, the interest in understanding and predicting an athlete's performance throughout his/her career is motivated by the need to evaluate the efficacy of training programs, anticipate fatigue to prevent injuries and detect unexpected of disproportionate increases in performance that might be indicative of doping. Moreover, fast evolving data gathering technologies require up to date modelling techniques that adapt to the distinctive features of sports data. In this work, we propose a hierarchical Bayesian model for describing and predicting the evolution of performance over time for shot put athletes. To account for seasonality and heterogeneity in recorded results, we rely both on a smooth functional contribution and on a linear mixed effect model with heteroskedastic errors to represent the athlete-specific trajectories. The resulting model provides an accurate description of the performance trajectories and helps specifying both the intra- and inter-seasonal variability of measurements. Further, the model allows for the prediction of athletes' performance in future seasons. We apply our model to an extensive real world data set on performance data of professional shot put athletes recorded at elite competitions.
	
\end{abstract}

\section{Introduction}
\label{intro}

\begin{figure*}
	\includegraphics[width = 1.0\textwidth]{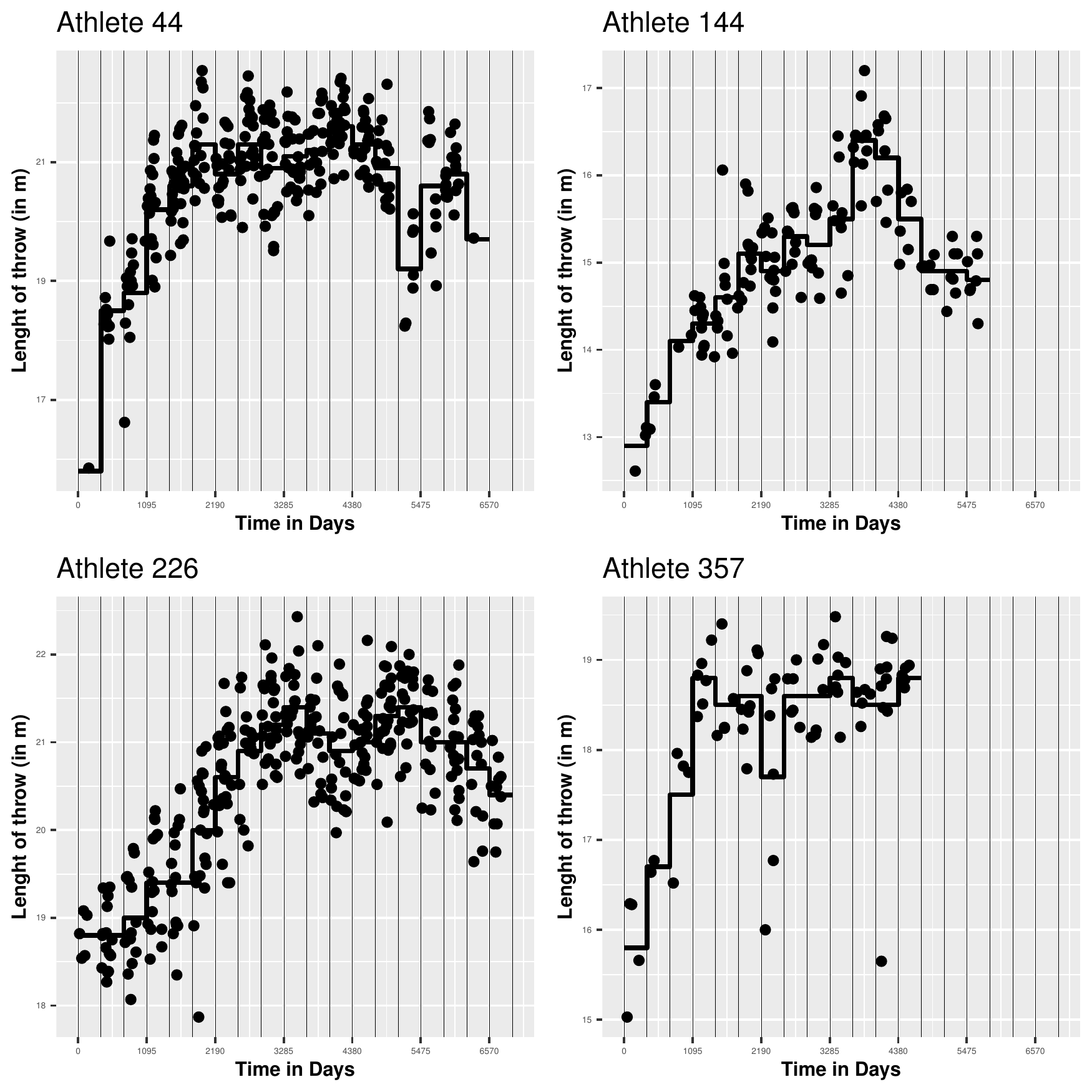}
	\caption{Each panel displays the performance results (points) of a professional shot put athlete throughout his/her career at elite competitions. Performance is measured in meters (length of the throw) and plotted against the days elapsed since January 1st of each athlete's career starting year. The dark vertical lines represent seasons changing points, namely new year days.}
	\label{data}
\end{figure*}

Shot put is a track and field event involving throwing (``putting") the shot, a metal ball (7.26kg/16lb for men, 4kg/8.8lb for women), with one hand as far as possible from a seven-foot diameter (2.135m) circle. In order for each put to be considered valid, the shot must not drop below the line of the athlete’s shoulders and must land inside a designated 35-degree sector. Athletes commonly put four to six times per competition, and their best performance is recorded. Figure \ref{data} displays the results of elite shot put competitions for four athletes with careers of different lengths. 
Furthermore, several athlete-specific variables (e.g., age, gender, doping history, nationality, etc.) that might be of interest in a sport analytics perspective are naturally gathered during elite competitions or can be easily retrieved retrospectively. Shot put competitions have been part of the modern Olympics since their revival, hence there exist complete and long-established data sets for shot put performances that can be exploited for performance analysis.  \\
\indent Sportive competitions might display some sort of seasonality in the results. We underline that here the term ``seasonality" is not used to indicate a cyclical behaviour of the results over time as in the literature of time series but, rather, a time dependent gathering of observations. On one hand competitions are traditionally concentrated in some months of the year, and on the other hand weather and environmental conditions may affect the performances or even the practicability of the sport itself. Straightforward examples are sports leagues that run on a different schedule and host their season openers, playoffs and championships at different times of the year or, even more dramatically, winter sports as opposed to outdoor water sports. Shot put events range over the whole year, with indoor competitions held during Winter months and major tournaments, like the Olympics, Diamond League and World Championship organised during Summer. Therefore, it is reasonable to say that seasons coincides with calendar years. In Figure \ref{data}, vertical lines represent new years' days: the time point at which seasons change. To provide an accurate representation of the data, seasonality effects need to be taken into account. Note that, from a modelling perspective, the amplitude of seasons is arbitrary, and ideally any model should be easily adapted to any sport with any type of seasonality. \\
\indent There exist a well established literature of both frequentist and Bayesian contributions to performance data analysis in various sports. We mention, among others, \cite{Tri} and \cite{Intro1} in triathlon, \cite{swim} and \cite{swim2} in swimming, and \cite{cricket} in cricket. \cite{batting} describe batting performance in baseball, \cite{deca} performance evolution in decathlon, \cite{golf} in golf, and \cite{Orani} in tennis. Further, \cite{aging} rely on a Bayesian latent variable model to investigate age-related changes across the complete lifespan of basketball athletes on the basis of exponential functions.\\
\indent In this work, we are interested in describing the evolution of performances of professional shot put athletes throughout their careers. To this end, we propose a Bayesian hierarchical model where results of each athlete are represented as error prone measurements of some underlying unknown function. The distinctive contribution of our approach is the form of such function, which has an additive structure with three components. First, we consider a smooth functional contribution for capturing the overall variability in athletes' performances. For this component, we follow the approach in \cite{M1}, who propose a Bayesian latent factor regression model for detecting the doping status of athletes given their shot put performance results and other covariates. However, the authors limit the analysis to data collected from 2012, whereas we aim at describing the trajectories in performance over the whole time span available for our data (1996 to 2016). For this time span, a global smoothness assumption for the trajectories could be too restrictive. Indeed, data may exhibit jumps localised at fixed and shared time points among all athletes. See, for example, athlete 303 in Figure \ref{data}, whose results show a jump between the second and fourth season (calendar year) of his/her career. The presence of jumps between seasons is even more striking when yearly average performances are considered. We highlight this behaviour by showing yearly averages via thick horizontal lines in Figure \ref{data}. Accordingly, the description of this yearly dependent contribution is delegated to a mixed effect model, that quantifies the seasonal mean for each athlete as a deviation from a grand mean. This component captures the inter-seasonal variability of the data set, whereas the smooth functional component describes the intra-seasonal evolution of performances. Finally, we complete our model specification accounting for the effect of a selection of time dependent covariates through a regressive component. We embed our model in a Bayesian framework by proposing suitable prior distributions for all parameters of interest. We believe the presented model represents a flexible tool to analyse evolution of performances in measurable sports, namely, all those disciplines for which results can be summarised by a unique measure (e.g., distance, time or weight).  \\
\indent The rest of the paper is organised as follows. In Section 2, we describe the motivating case study. In Section 3, we present the proposed model and briefly discuss possible alternative settings. Moreover, we elicit priors for our Bayesian approach. In Section 4, we outline the algorithm for posterior computation. In Section 5, we discuss posterior estimates, argue on the performance of the model and interpret the model's parameters form a sports analytic perspective. Conclusions are presented in Section 6. Finally, the Appendix includes the complete description of the MCMC algorithm. 

\section{The World Athletics shot put data set}
\label{dataset}

\begin{figure*}
\centering
	\includegraphics[width=\textwidth]{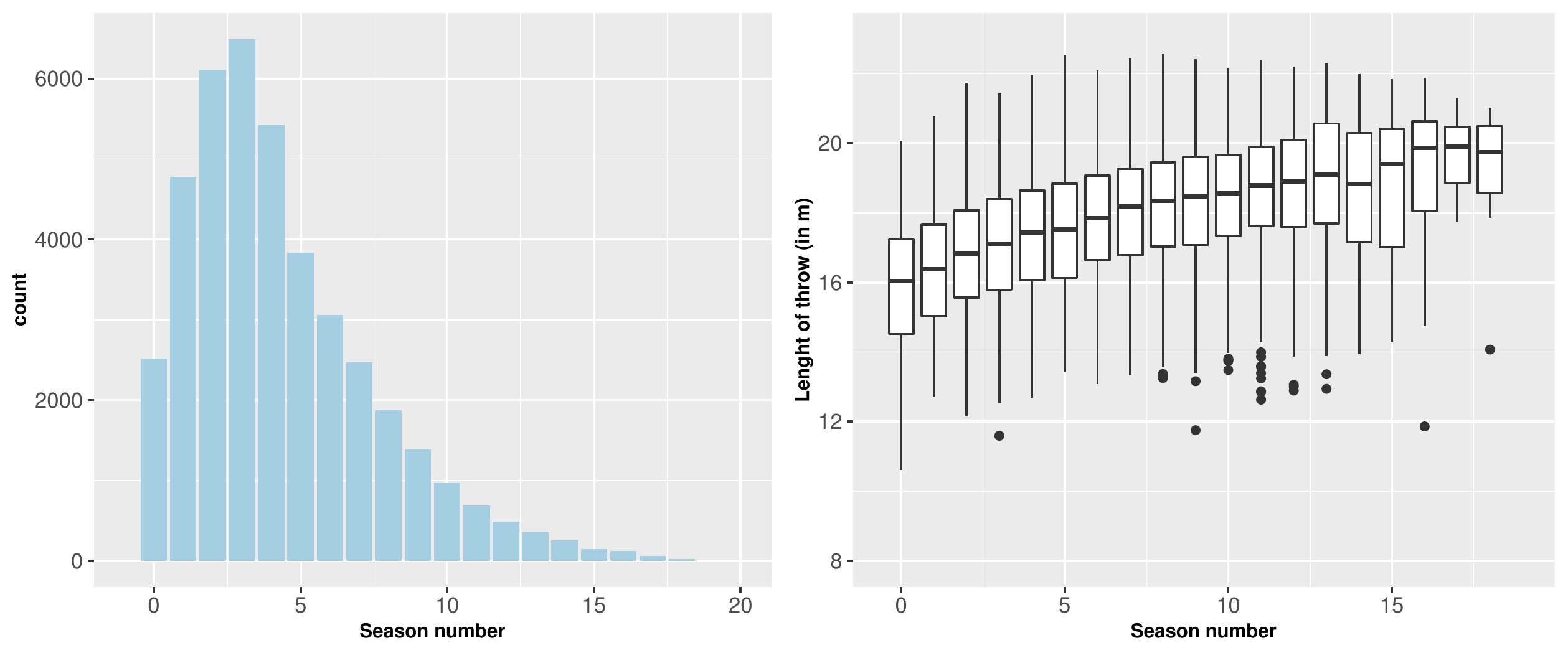}
	\caption{Left: total number of athletes per season. Right: each boxplot shows the distribution of the athletes' mean performances within each season.}
	\label{seasons}
\end{figure*}

\begin{figure}
	\centering
	\includegraphics[width=\textwidth]{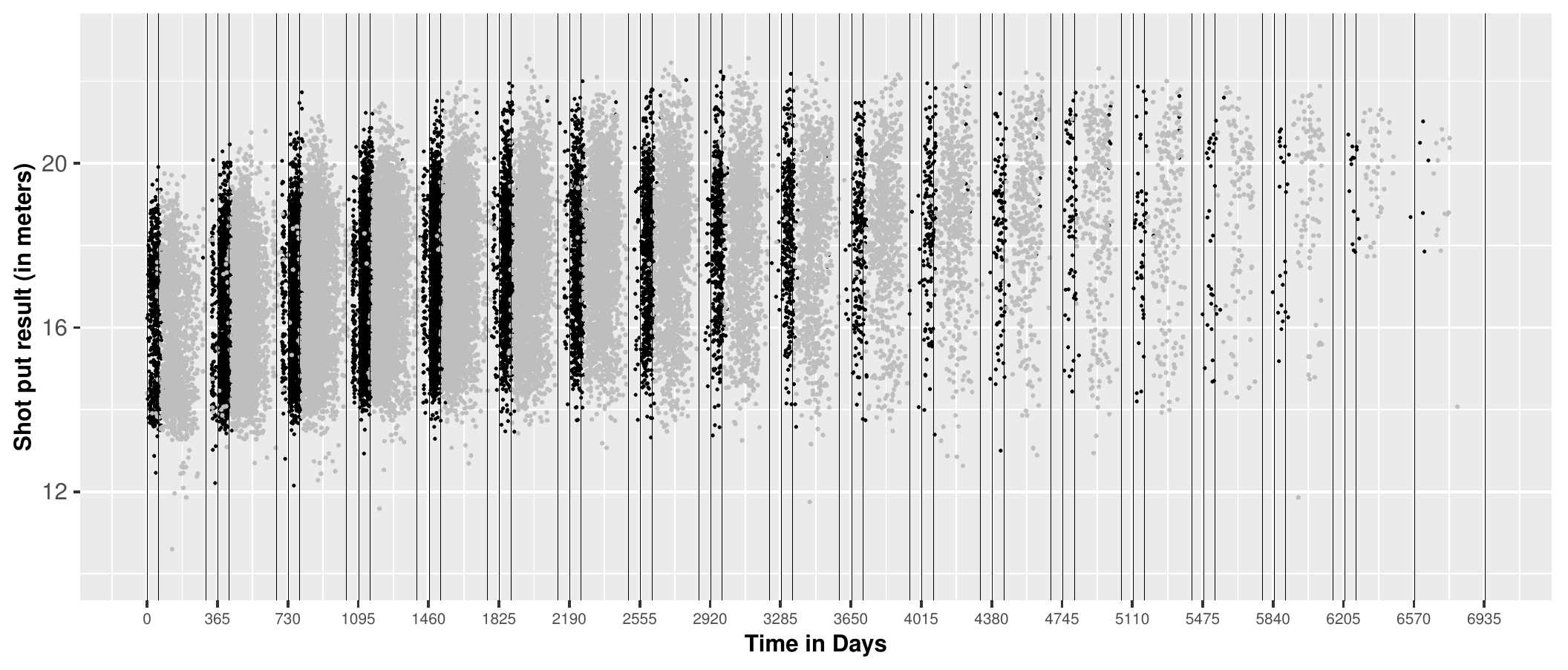}
	\caption{Performance results displayed according to the variable environment, which takes values indoor (black) and outdoor (gray). Vertical lines corresponding to the label ticks represent season changes, whereas the two enclosing it indicate the boundaries of winter months.}
	\label{Environment}
\end{figure}

World Athletics (WA) is the world governing body for track and field athletic sports. It provides standardized rules, competition programs, regulated technical equipment, a list of official world records and verified measurements. The data at our disposal was obtained with permission from an open results database (\url{www.tilastopaja.eu}) following institutional ethical approval (Prop\_72\_2017\_18). The data set comprises 56,000 measurements of WA recognized elite shot put competitions for 1,115 athletes from 1976 to 2016. For each athlete, the data set reports the date of the event, the best result in meters, the finishing position, an indication of any doping violation during the athlete's career as well as demographic information (athlete's name, WA ID number, date of birth, sex and country of birth). \\
\indent In this work, we restrict our analysis to results for athletes performing after 1996. Indeed, we pursue consistency of measurement accuracy, and 1996 represents a turning point in anti-doping regulation and fraud detection procedures. The resulting data set is still sufficiently broad for our purposes. It contains 41,033 observations for 653 athletes (309 males and 344 females). The outcome of interest is the shot distance, which ranges from a minimum if $10.6$ up to a maximum if $22.56$ meters, with a mean of $17.30$ meters. \\
\indent As shown in Figure \ref{data} for a selection of athletes, data are collected over time. Hereafter we will denote as $t_{ij}$ the time at which the $j$-th observation for athlete $i$ is recorded. $t_{ij}$ corresponds to the time elapsed from January 1st of each athlete's career starting year to the date of the competition. Accordingly, equal time values for different athletes are likely to specify distinct years, but the same moment in those athletes' careers. Moreover, different athletes will have observations ranging over a large time span, according to the length of their careers. Having described seasons as calendar years, athletes will also compete in a different number of seasons. Figure \ref{seasons} shows the number of athletes per season as well as boxplots of the distribution of their mean performances across the various seasons. Athletes with the longest careers have been playing for 19 years. A general increasing trend in performance can be observed as a function of career length (right panel in Figure \ref{seasons}) or, equivalently, the age of the athlete. In the following, we will discuss two different modelling choices for age, respectively, accounting for its time dependence and considering age as a fixed quantity, namely the age of the athlete at the beginning of his/her career. \\
\indent Table 1 reports descriptive statistics suggesting how sex, environment and doping have an effect on the average value of the result. We point out that in our dataset we only have 18 athletes who tested positive for doping at some point in their career. Information on the date the test was taken (or if multiple tests were taken) is not available in the data. As expected, performances for men are, on average, higher than for women. Similar effects, despite less evident in magnitude, also hold true for the variable environment, which takes values indoor and outdoor. Regarding environment, a further remark is in order. We already pointed out that major WA events take place outdoor (27,800 observations) during Summer months, whereas less competitive events are held inside between November and March (13,200 observations). Figure \ref{Environment} displays results in gray when recorded outdoor, and in black otherwise. We can clearly see that field events gather in Summer months, whereas indoor events take place during Winter (solid lines).

\begin{table}
\centering
	\caption{Performance results conditioned on covariates.}
	\label{result}
	\begin{tabular}{lllll}
		\toprule
		& Mean & Sd & Max & Min \\
		\midrule
		Total & 17.30 & 1.78 & 22.56 & 10.6 \\
		Women & 16.09 & 1.35 & 21.70 & 10.6 \\
		Men & 18.55 & 1.21 & 22.56 & 12.93 \\
		Not Doped & 17.30 & 1.79 & 22.56 & 10.6 \\
		Doped & 17.77 & 1.35 & 20.88 & 13.55 \\
		Indoor & 17.17 & 1.70 & 22.23 & 12.15 \\
		Outdoor & 17.38 & 1.81 & 22.56 & 10.6 \\
		\bottomrule
	\end{tabular}
\end{table}

\section{The model}
\label{baymodel}

Let $n$ denote the total number of athletes in the study. We assume that shot put performances for athlete $i$ are given by noisy measurements of an underlying function $g_i(t_{ij})$: 
\begin{equation}
	\label{full}
	y_{ij}=g_i(t_{ij}) + \epsilon_{ij}
\end{equation}
with $\epsilon_{ij} \stackrel{iid}{\sim} N(0, \psi^2)$ independent errors. Recall $t_{ij}$ is the time at which the $j$-th observation for athlete $i$ is collected, for $j = 1, \ldots, n_i$, where $n_i$ is the total number of measurements available on athlete $i$.\\
We further suggest an explicit functional form for $g_i(t_{ij})$:
\begin{equation}
	g_i(t_{ij}) = f_i(t_{ij}) + \mu_{is} + \boldsymbol{x}_{i}(t_{ij}) \boldsymbol{\beta}
	\label{model}
\end{equation}
where $f_i(t)$ is a smooth functional component for intra-seasonal variability, $\mu_{is}$ a season-specific intercept, and $\boldsymbol{x}_{i}(t) \boldsymbol{\beta}$ is an additional multiple regression component. Here $s \in \{ 1,2, \ldots, S_i \}$ indicates the season in which the shot was recorded. Specifically, $\mu_{is} \equiv \mu_i(t_{ij}) = \sum_{s=1}^{S_i} \mu_{is} \ \mathbb{I}_{(t_i^s, t_i^{s+1})}(t_{ij})$ is an athlete-specific step function taking value $\mu_{is}$ for all time points in season $s$, delimited by $t_i^s$ and $t_i^{s+1}$. For a complete treatment of the notation used insofar, please refer to Table \ref{notation}. \\ 
\indent We will now discuss each of the three terms in Eq.~\eqref{model} in more detail. 

\begin{table}
\centering
	\caption{Mathematical notation}
	\label{notation}
	\begin{tabular}{ll}
		\toprule
		Symbol & Meaning   \\
		\midrule
		\textit{i} & Index identifying the athlete \\
		\textit{j} & Index identifying a specific observation \\
		\textit{n} & Total number of athletes \\
		\textit{N} & Total number of observations \\
		$t_{ij}$ & Time point at which the \textit{j}'th observation of athlete \textit{i} is recorded\\
		$S_i$ & Total number of seasons for athlete $i$\\
		\textit{s} & The currently considered season\\
		$g_{is}$ & Number of observations in season $s$ for athlete $i$ \\
		\textit{r} & The number of covariates to be considered\\
		$y_{ij}$ & Response variable at time $j$ for athlete $i$ \\
		\bottomrule
	\end{tabular}
\end{table}

\subsection{The functional component}
\label{funpart}
The functional component $f_i(t)$ is meant to capture the subject-specific global evolution of the response variable. It explains the global dependence of the data from time. We require that these functions display a smooth behaviour: the latter is assured by assuming $\{ f_i (t)\}_{i=1}^n $ are linear combinations of smooth basis functions, $\{b_m (t)\}_{m=1}^p$. Note that, both the nature and the number $p$ of these bases are to be determined according to some properties we wish them to satisfy. In particular, we assume: 
\begin{equation}
	f_i(t) = \sum_{{m}=1}^p \theta_{im} b_m(t)
	\label{fmodel}
\end{equation}
where $\{b_m(t) \}_{m=1}^p$ represent the B-spline basis \citep{boor} and $\{ \theta_{im} \}_{m=1}^p$ are subject-specific coefficients.\\
\indent We briefly recall that the B-spline basis of degree $k$ on $[L, U]$ is a collection of $p$ polynomials defined recursively on a sequence of points, known as knots, and indicated with $L \equiv t_1 \leq \ldots \leq t_{p+k+1} \equiv U$. 
We follow the common approach of choosing $k = 3$, leading to cubic splines \citep[see, for instance,][]{cubic}. Moreover, we assume the knot sequence to be equispaced and $(k+1)$-open. That is, the first and last $k+1$ knots are identified with the extremes of the definition interval, whereas the remaining $p-k-1$ knots divide said interval into sets of the same length. Under these assumptions, each basis function $b_j(t)$ has compact support over $k+1$ knots, precisely $[t_j, t_{j+k+1}]$. Moreover, together they span the space of piecewise polynomial functions of degree $k$ on $[L, U]$ with breakpoints $\{ t_n \}_{n=1}^{p+K+1}$. Finally, such functions are twice continuously differentiable at the breakpoints, de facto eliminating any visible type of discontinuity and providing a smooth result.\\
\indent The number of basis functions is chosen to be large enough for sufficiently many basis functions to have a completely enclosed support in a given season. In particular, rescaling time to $[0,1]$ (for computational purposes and easier prior definition) seasons have amplitude $0.054$ and, if we want at least a local basis to be completely supported in one of them, we need $4$ knots to fall in it. Accordingly, we require about 75 internal knots and 80 degrees of freedom. \\
\indent With the sake of tractability, a low dimensional representation of the individual curves is of interest. Following the approach by \cite{M1}, we exploit a sparse latent factor model on the basis coefficients:
\begin{equation}
	\theta_{im} = \sum_{l=1}^k \lambda_{ml} \eta_{il} + \xi_{im}
	\label{factors}
\end{equation}
where $\lambda_{ml}$ are the entries of a $(p \times k)$ factor loading matrix $\boldsymbol{\Lambda}$, and $\boldsymbol{\eta_{i}}$ is a vector of $k$ latent factors for subject $i$. Finally, $\boldsymbol{\xi_i} = ( \xi_{i1}, \ldots , \xi_{ip} )$ is a residual vector, independent of all other variables in the model. We assume:
\begin{equation}
	\boldsymbol{\eta_i} \stackrel{iid}{\sim} N_k(\boldsymbol{0}, \boldsymbol{I})
\end{equation}
and the error terms $\boldsymbol{\xi_i}$ are assumed to have normal distribution with diagonal covariance matrix, $\boldsymbol{\xi_i} \stackrel{iid}{\sim} N_p(\boldsymbol{0},  diag(\sigma_1^{-2}, \ldots, \sigma_p^{-2}))$, with $\sigma_j^{-2} \stackrel{iid}{\sim} Ga(a_{\sigma}, b_{\sigma})$.\\
\indent For the modelling of the factor loading matrix $\boldsymbol{\Lambda}$, we follow the approach in \cite{shrink}. Specifically, we adopt a multiplicative gamma process shrinkage prior which favours an unknown but small number of factors $k$. The prior is specified as follows:
\begin{align}
	\label{functional}
	& \lambda_{ml} | \phi_{ml}^{-1},  \tau_l^{-1} \stackrel{iid}{\sim} N\left(0, \phi_{ml}^{-1} \tau_l^{-1}\right) \qquad  \text{with} \\
	&  \phi_{ml} \sim  Ga \biggl( \frac{\nu_{\phi}}{2},\frac{\nu_{\phi}}{2} \biggr) \qquad    \tau_l=\prod_{v=1}^{h} \varpi_v  \nonumber \\
	&  \varpi_1 \sim Ga(a_1,1)    \qquad 
	\varpi_v \sim Ga(a_v,1) \nonumber
\end{align} 
Here $\tau_l$ is a global stochastically increasing shrinkage parameter for the $l$-th column which favors more shrinkage as the column index increases. Similarly, $\phi_{ml}$ are local shrinkage parameters for the elements in the $l$-th column. 
\cite{shrink} describe a procedure to select the number of factors $k \ll p$ adaptively. We follow their lead, thus $k$ is not set a priori but automatically tuned as the Gibbs sampler progresses. Refer to \cite{shrink} for details. Note that, by combining Eq. \eqref{fmodel} and Eq. \eqref{factors} one gets:
\begin{equation*}
	f_i(t) = \sum_{l=1}^k \eta_{il} \Phi_{l}(t) + r_i(t)
\end{equation*}
where $\Phi_{l}(t) = \sum_{m=1}^p \lambda_{ml}  b_m(t)$ is a new, unknown non-local basis function learnt from the data and $r_i(t)$ a functional error term. Although the number of pre-specified basis function $p$ is potentially large, the number of ``operative'' bases $k$ is $k \ll p$ and learnt from the data. \\

\subsection{The seasonal component}
\label{seaspart}

Early graphical displays and straightforward exploratory analysis suggest a significant variability of the average response across seasons as displayed in Figure \ref{data}. Namely, performances prove to be gathered over predetermined time intervals, the seasons (calendar years). However, it is reasonable to expect some degree of dependence for the average performance across seasons. To model such dependence, an autoregressive model for seasonal intercepts can be proposed. The idea behind this choice is to allow for borrowing of information across seasons, in the sense that the seasonal intercept $\mu_{is}$ at season $s$ is influenced by the intercept at season $s-1$ through the autoregressive coefficient $\rho_i$. Namely,
\begin{equation}
	\mu_{is} \ | \ \rho_i, \sigma_{\mu}^2 \stackrel{iid}{\sim}  N\left(\rho_i \mu_{i(s-1)}, \sigma_{\mu}^2\right)
	\label{autoregress}
\end{equation}
However, when we first implemented this model, we noted how residuals presented a pattern which we would like to intercept with a finer model. Therefore, we consider a random intercept model with Normal Generalized Autoregressive Conditional Heteroskedastic (GARCH) errors \citep{GARCH}. Specifically,
\begin{gather}
	\mu_{is} \ | \ m, h_{is} = m + \zeta_{is} \stackrel{iid}{\sim}  N(m, h_{is})\\
	h_{is} = \alpha_0 + \alpha_1 \zeta_{is-1}^2 + \varpi h_{is-1} \label{garch}
\end{gather}
where $\alpha_0 > 0, \alpha_1 \geq 0$ and $\varpi \geq 0$ to ensure a positive conditional variance and $\zeta_{is} = \mu_{is} - m$ with $ h_{i0} = \zeta_{i0} := 0 $ for convenience. The additional assumption of wide-sense stationarity with 
\begin{gather*}
	\mathbb{E}(\zeta_t)=0\\
	\mathbb{V}ar(\zeta _{t}) = \alpha_{0} (1- \alpha_{1} - \varpi)^{-1} \\
	\mathbb{C}ov(\zeta _{t}, \zeta _{s}) = 0 \text{ for }  t \neq s
\end{gather*}
is guaranteed by requiring $\alpha_{1} + \varpi < 1$, as proven by \cite{GARCH}.\\
\indent Three parameters of the seasonal component require prior specification: the overall mean $m$ and the conditional variance parameters, $\varpi$ and $\boldsymbol{\alpha}=(\alpha_0, \alpha_1)^\top$. For the autoregressive and heteroskedastic parameters of the GARCH model, we propose non-informative priors satisfying the positivity constraint. For the overall mean parameter, we rely on a more informative Normal prior centered around the mean suggested by posterior analysis of preliminary versions of the model. In particular:

\begin{gather}
	\nonumber m \sim N(\mu_{m_0}, \Sigma_{m_0}) \\ 
	\boldsymbol{\alpha} \sim  N_2(\mu_{\alpha}, \Sigma_{\alpha}) \ \mathbb{I} \{ \boldsymbol{\alpha} > 0 \} \label{GARCHprior} \\
	\nonumber \varpi \sim  N(\mu_{\varpi}, \Sigma_{\varpi}) \ \mathbb{I} \{ \varpi \geq 0 \} 
\end{gather}
where $\boldsymbol{\alpha} = (\alpha_0,\alpha_1)$ is a bidimensional vector. We complete the model specification assuming that the parameters are statistically independent and noticing that the hypothesis needed for wide-sense stationarity do not translate into actual prior conditions on the parameters. Hence, one of the objects of our analysis becomes to test whether the constraint  $\alpha_{1} + \varpi < 1$ holds true.

\subsection{Covariates}
\label{covpart}

We consider the effect of three covariates, gender, age and environment, and assume conjugate prior choices for the covariates coefficients:
\begin{gather}
	\nonumber \boldsymbol{\beta} \stackrel{iid}{\sim} N(\boldsymbol{\boldsymbol{\beta}_0}, \sigma_{\beta}^{2} \boldsymbol{\mathbb{I}}) \\
	\sigma_{\beta}^{-2} \sim Ga \biggl( \frac{\nu_{\beta}}{2},\frac{\nu_{\beta} \sigma_{\beta}^2}{2} \biggr)
	\label{reg}
\end{gather}

\section{The Bayesian update}
\label{Update}
 Because of the additive nature of the overall sampling model \eqref{full} - \eqref{model}, we are able to exploit a blocked Gibbs sampler grouping together the parameters of the three modelling components described in Section \ref{baymodel}.
Note first that, because of the high dimensionality of the problem, it is computationally convenient to choose conditionally conjugate prior distributions for the parameters. Indeed, conjugacy guarantees analytical tractability of posterior distributions. In some cases, specifically for the conditional variances of GARCH errors, no conjugate model exists and updates rely on an adaptive version of the Metropolis Hastings algorithm for posterior sampling. \\
\indent Algorithm \ref{Gibbs} outlines our sampling scheme, while details are presented in Appendix~\ref{appendix}. As far as the parameters of the functional component $\boldsymbol{\theta}_i$ are concerned, we follow \cite{M1} by choosing conditionally conjugate prior distributions so that the update proceeds via simple Gibbs sampling steps. Analogously, the update of the regression coefficients $\boldsymbol{\beta}$ and the error term $\psi$ proceeds straightforwardly by sampling from their full conditional posterior distributions. Conjugate priors for the GARCH parameters $m, \varpi$ and $\boldsymbol{\alpha}$ are not available, therefore we resort to adaptive Metropolis schemes to draw values from their full conditionals. Specifically, we build an adaptive scale Metropolis such that the covariance matrix of the proposal density adapts at each iteration to achieve an \emph{optimal} acceptance rate \citep[see][]{HAARIO}. \\
\indent Further details about the algorithm can be found in the Appendix \ref{appendix}, whereas code is available at \url{https://github.com/PatricDolmeta/Bayesian-GARCH-Modeling-of-Functional-Sports-Data.}  

\IncMargin{5mm}
\begin{algorithm}[t]
	\label{Gibbs}		  
	\vspace{2mm}
	\hspace{-5mm} \KwData{$y_{ij} = (y_{11},\ldots,y_{nn_n})$}
	\hspace{-5mm}  Set the required MCMC sample size $G$, the burn-in period $g_0$ and the thinning parameter $g_s$. \\[0mm]
	
	\SetKwBlock{Begin}{Initialise}{end}
	\Begin{ $\boldsymbol{\theta}_i^{(0)}, \boldsymbol{\mu}_i^{(0)}, m^{(0)},\varpi^{(0)},\boldsymbol{\alpha}^{(0)}, \boldsymbol{\beta}_i^{(0)}, \psi_i^{(0)} $\\
	}
	
	\SetKwBlock{Begin}{For $g=0,\ldots,G$}{end}
	\Begin{
		\SetKwBlock{Begin}{Update functional component}{end}
		\Begin{
			Set partial residuals $y_{ij}^{(1)(g)} = y_{ij} - \mu_{i,s}^{(g)} - \boldsymbol{x_i}(t_{ij}) \boldsymbol{\beta}^{(g)}$ \\
			Update $\boldsymbol{\theta}_i^{(g+1)}$ on the base of Appendix A.1
		}
		
		\SetKwBlock{Begin}{Update seasonal component}{end}
		\Begin{
			Set partial residuals $y_{ij}^{(2)(g)} = y_{ij} - f_i(t_{ij})^{(g)} - \boldsymbol{x_i}(t_{ij}) \boldsymbol{\beta}^{(g)}$ \\
			Update $\boldsymbol{\mu}_i^{(g+1)}, m^{(g+1)},\varpi^{(g+1)},\boldsymbol{\alpha}^{(g+1)} $ on the base of Appendix A.2
		}
		
		\SetKwBlock{Begin}{Update regressive component}{end}
		\Begin{
			Set partial residuals $y_{ij}^{(3)(g)} = y_{ij} - f_i(t_{ij})^{(g)} - \mu_{i,s}^{(g)} $ \\
			Update $\boldsymbol{\beta}_i^{(g+1)}$ on the base of Appendix A.3
		}
		
		\SetKwBlock{Begin}{Update error term}{end}
		\Begin{
			Set partial residuals $\epsilon_{ij}^{(g)} = y_{ij} - f_i(t_{ij})^{(g)} - \mu_{i,s}^{(g)} - \boldsymbol{x_i}(t_{ij}) \boldsymbol{\beta}^{(g)}$ \\
			Update $\psi^{(g+1)}$ on the base of Appendix A.4
		}
		
		\textbf{Return} $\boldsymbol{\theta}_i^{(g)}, \boldsymbol{\mu}_i^{(g)}, m^{(g)},\varpi^{(g)},\boldsymbol{\alpha}^{(g)}, \boldsymbol{\beta}_i^{(g)}, \psi_i^{(g)} $ for $g = g_0, g_0+g_s, g_0 + 2g_s, \ldots,G$
	}
	\caption{Gibbs Sampler}
	
\end{algorithm}

\section{Posterior analysis}
\label{estimation}

The idea of estimating trajectories for athletes' performances is a natural pursuit for the model specification we adopted. Indeed, describing observations as error prone measurements of an unknown underlying function suggests evaluating such function, once retrieved, on any number of points of interest. In practice, we will generate a fine grid of $T$ equispaced time points: $\{ t_k \}_{k=1}^T$  between $0 \equiv t_1$ and $1 \equiv t_T$ and evaluate the function on this grid. \\
\indent In particular, we start by evaluating the athlete-specific functional component by exploiting the basis function representation. Being:
\[ \boldsymbol{\Theta}_i = \begin{bmatrix}
	\theta_{i1}^{(1)} &\theta_{i2}^{(1)} & \ldots & \theta_{ip}^{(1)} \\
	\theta_{i1}^{(2)} &\theta_{i2}^{(2)} & \ldots & \theta_{ip}^{(2)} \\
	\vdots & \vdots & \ddots & \vdots \\
	\theta_{i1}^{(G)} &\theta_{i2}^{(G)} & \ldots & \theta_{ip}^{(G)} 
\end{bmatrix}
\]
the matrix of individual-specific spline basis coefficients for all iterations $g = 1, \ldots G$ and \[ \boldsymbol{b}^\top = \begin{bmatrix}
	b_{1} (t_1) &b_{1} (t_2) & \ldots & b_{1} (t_k) &  \ldots & b_{1}(t_T) \\
	b_{2} (t_1) &b_{2} (t_2) & \ldots & b_{2} (t_k) &  \ldots & b_{2}(t_T) \\
	\vdots & \vdots & \ddots & \vdots & \ddots & \vdots \\
	b_{p} (t_1) &b_{p} (t_2) & \ldots & b_{p} (t_k) &  \ldots & b_{p}(t_T) 
\end{bmatrix}
\]
all values of a $p$-dimensional, degree-$3$, spline basis on a set of $T+1$ equispaced knots in the unit interval, the estimated contribution of the functional component to the overall trajectory is, at each iteration: 
\[ {f}_i^{(g)}(t) = \sum_{m=1}^p {\theta}_{im}^{(g)} b_m(t) = {\Theta}_i^{(g)} \boldsymbol{b}_{t}^\top \quad \text{ for } t = {t_1, \ldots, t_T},
\]
where ${\Theta}_i^{(g)}$ corresponds to the $i$-th row of matrix $\boldsymbol{\Theta}_i$. \\
\indent As for the seasonal linear mixed effect, we modelled it as a piecewise continuous function taking individual- and season-specific values. Hence, when retrieving its estimated effect on any point in the time grid, we need to determine which season it belongs to. As discussed in Section 3, time is rescaled so that equal values across individuals indicate the same day of the year, possibly in different years. Therefore, season changes, that occur at new year's days, can be easily computed by straightforward proportions. At this point, the season to which $t_k$ belongs to is obtained by comparison with the season thresholds. In the following Equation, the indicator variable $ \chi_{(t \in s)}$ determines to which season each time point belongs to. Accordingly, the estimated contribution of the seasonal component to the overall trajectory is, at each iteration: 
\[ {\mu}_i^{(g)}(t) = \sum_{s=1}^{S_i} {\mu}_{is}^{(g)} \chi_{(t \in s)} \quad \text{ for } t = {t_1, \ldots, t_T}.
\]
\indent Lastly, the regressive component has to be taken into account. The estimated contribution of the regressive component to the overall trajectory is, at each iteration: 
\[ \sum_{l=1}^{r} x_{il} (t) {\beta}_{l}^{(g)} = \boldsymbol{x}_{i}(t) {\boldsymbol{\beta}}^{(g)}  \quad \text{ for } t = {t_1, \ldots, t_T}.
\]
\indent Given the three components, the overall estimate of the underlying function is obtained by adding these three components. In particular, the estimated mean trajectory can be written as:
\begin{equation}
	\widehat{y_i(t)} = \frac{1}{G}\sum_{g=1}^G  {f}_i^{(g)}(t) +  {\mu}_i^{(g)}(t) + \boldsymbol{x}_{i}(t) {\boldsymbol{\beta}}^{(g)} \quad \text{ for } t = {t_1, \ldots, t_T} \label{compll}
\end{equation}
\indent Similarly, $95 \%$ credible intervals can be computed to quantify uncertainty around our point estimate.

\subsection{Model application}

In this Section, we fit different specifications of our model to the data described in Section \ref{dataset}.\\
\indent In general, we consider the additive structure of the sampling model illustrated in Equation \ref{model}. Table \ref{modelsummary} reports an overview on of the six models we compare. In Model $M_1$, the B-sline basis functions have 80 degrees of freedom, the seasonal component has GARCH errors and three regressors are taken into account: \textit{sex, age and environment}. $M_2$ represents a slight modification of $M_1$ given by the fixed-age implementation. Here we consider the covariate \textit{age} not as a time dependent variable, but as a fixed value given by the age at the beginning of each athlete's career. In model $M_3$ a simpler dependence structure among the seasonal effects is used. Namely, we assume an autoregressive model for $\mu_is$ (see Eq. \ref{autoregress}). For model $M_4$, we simply consider a larger number of basis functions, i.e. 120, accounting for up to three splines having support in a season and hence meant to better capture the intra-seasonal variability. Finally, models $M_5$ and $M_6$ allow for \textit{doping} as additional covariate, both in the case of the time-dependent and time-independent specification of \textit{age}.\\
\indent Priors were chosen as discussed in Section \ref{baymodel}, and with hyperparameter choices summarized in Table \ref{hyper}. To argue on the choice of the informative prior for the overall mean parameter $m$, in Table \ref{comparison} we also report the results under a slight modification of model $M_1$, that we denote $M_1^{(2)}$, yielding a vague prior for $m$.

\begin{table}
\centering
	\caption{Models name and description.}
	\label{modelsummary}
	\begin{tabular}{ll}
		\toprule
		Symbol & Meaning   \\
		\midrule
		$M_1$ & 80 df B-splines, GARCH, covariates: sex, age (time dependent), env. \\
		$M_2$ & 80 df B-splines, GARCH, covariates: sex, age (time constant), env. \\
		$M_3$ & 80 df B-splines, AR, covariates: sex, age (t. dep.), env. \\
		$M_4$ & 120 df B-splines, GARCH, covariates: sex, age (t. dep.), env.\\
		$M_5$ & 80 df B-splines, GARCH, covariates: sex, age (t. dep.), env., doping\\
		$M_6$ & 80 df B-splines, GARCH, covariates: sex, age (t. const.), env., doping \\
		\bottomrule
	\end{tabular}
\end{table}

\begin{table}
\centering
	\caption{Hyperparameter choices. In the first column, we refer to the Equation where the hyperparameter first appears.}
	\label{hyper}
	\begin{tabular}{llll}
		\toprule
		Ref. & Hyp. & Value & Description    \\
		\midrule
		(\ref{factors}) & $a_{\sigma}$ & 1.0 & $1^{st}$ Gamma coeff. of error term in the factor exp. \\
		(\ref{factors}) & $b_{\sigma}$ & 0.3 & $2^{nd}$ Gamma coeff. of error term in the factor exp. \\
		(\ref{functional}) & $\nu_{\phi}$ & 9 & Gamma coeff.s of local shrink. param. $\phi_{ml}$ \\
		(\ref{functional}) & $a_{1}$ & 2.1 & $1^{st}$ Gamma coeff. of the $1^{st}$ global shrink. factor $\delta_1$ \\
		(\ref{functional}) & $b_{1}$ & 1.0 & $2^{nd}$ Gamma coeff. of the $1^{st}$ global shrink. factor $\delta_1$ \\
		(\ref{functional}) & $a_{l}$ & 2.1 & $1^{st}$ Gamma coeff. of the $l$-th global shrink. factor $\delta_l$ \\
		(\ref{functional}) & $b_{l}$ & 1.0 & $2^{nd}$ Gamma coeff. of the $l$-th global shrink. factor $\delta_l$ \\
		(\ref{GARCHprior}) & $\mu_{m_0}$ & -0.2 & Mean of the overall mean $m$ \\
		(\ref{GARCHprior}) & $\Sigma_{m_0}$ & 0.0001 & Variance of the overall mean $m$ \\
		(\ref{GARCHprior}) & $ \mu_{\alpha}$ & (0.0, 0.0) & Mean vector of the $\boldsymbol{\alpha}$ GARCH coeff. \\
		(\ref{GARCHprior}) & $ \Sigma_{\alpha} $ & $\mathbb{I}_2$ & Covariance matrix of the $\boldsymbol{\alpha}$ GARCH coeff. \\
		(\ref{GARCHprior}) & $ \mu_{\varpi}$ & 0.0 & Mean of the $\varpi$ GARCH coeff. \\
		(\ref{GARCHprior}) & $ \Sigma_{\varpi} $ & 1 & Variance of the $\varpi$ GARCH coeff. \\
		(\ref{reg}) & $\nu_{\beta}$ & 0.5 & $1^{st}$ Gamma coeff.s regression param. \\
		(\ref{reg}) & $\sigma_{\beta}$ & 0.5 & $2^{nd}$ Gamma coeff.s of regression param. \\
		(\ref{full}) & $\mu_{\psi}$ & 1.0 & Mean of the error variance $\psi$ \\
		(\ref{full}) & $\sigma_{\psi}$ & 1.0 & Variannce of the error variance $\psi$ \\
		\bottomrule
	\end{tabular}
\end{table}

For all experiments, inference is obtained via posterior samples drawn by the Gibbs sampler introduced in Section \ref{Update}. In particular, we ran $20,000$ iterations with a burn-in period of $60 \%$ and a thinning of $5$. Performances are compared by means of the logarithm of the pseudo marginal likelihood (LPML) index \citep{GEI}. This estimator for the log marginal likelihood is based on conditional predictive densities and provides an overall comparison of model fit, with higher values denoting better performing models. 

\begin{table}
	\centering
	\caption{Model and hyperparameter comparison for the models in Table \ref{modelsummary}.}
	\begin{tabular}{ll}
		\toprule
		Model & LPML \\ [0.5ex] 
		\midrule
		$M_1$ & $\boldsymbol{-45943}$  \\ 
		$M_1^{(2)}$ & -46573 \\
		$M_2$ & $\boldsymbol{-45472}$ \\
		$M_3$ & -46544  \\ 
		$M_4$ & -46314 \\
		$M_5$ & -48565 \\
		$M_6$ & -48122 \\
		\bottomrule
	\end{tabular}
	\label{comparison}
\end{table}

\begin{figure}
	\centering
	\includegraphics[width = 1.0\textwidth]{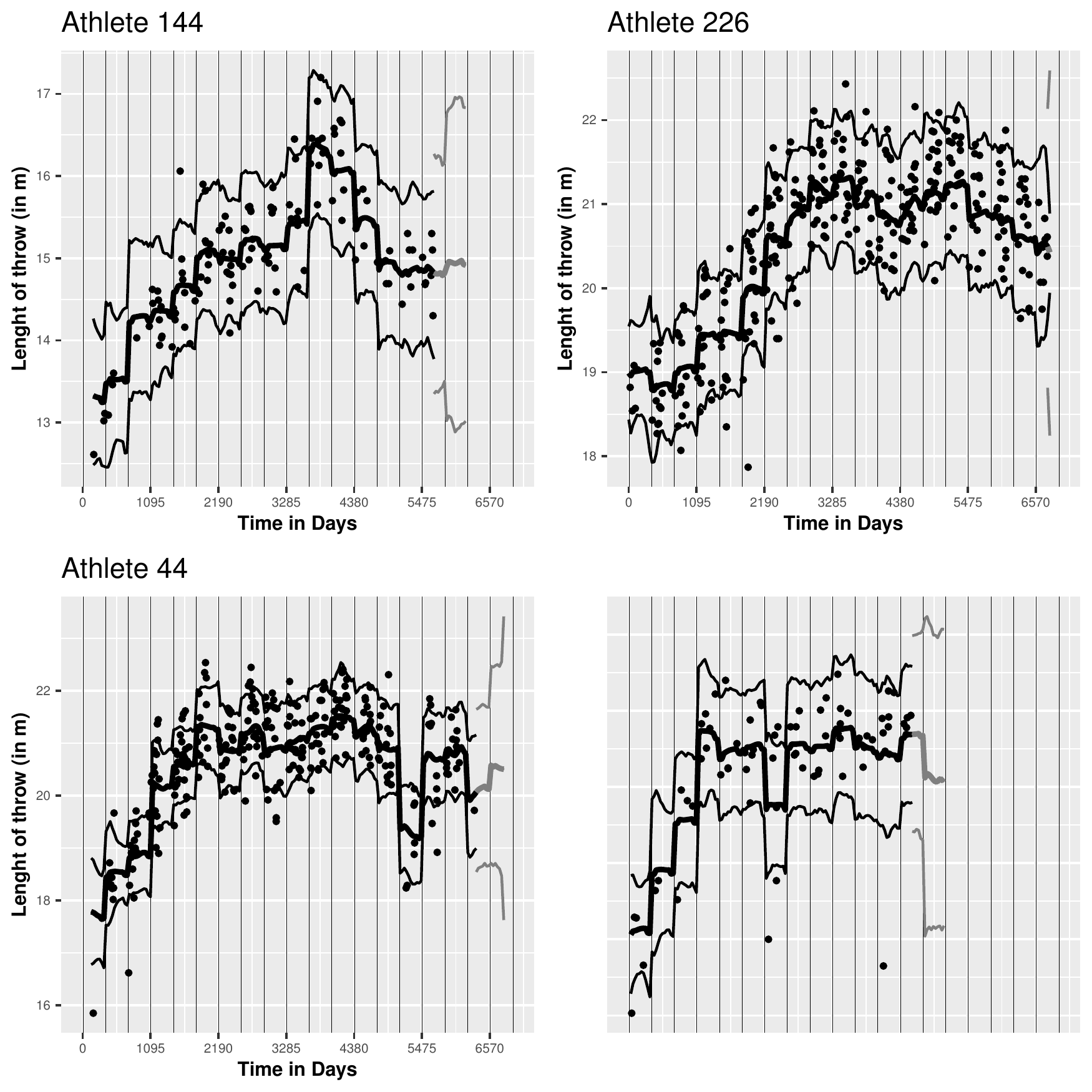}
	\caption{Performance trajectory estimates for a random selection of athletes. The $x$-axis denotes the time measured in days from January 1st of the first season of career, whereas on the y-axis there is the length of throw in meters. Vertical lines represent calendar years (seasons in our notation). The final part of each trajectory (grey) for which no observations are available, represents one-season-ahead performance prediction.}
	\label{Athl}
\end{figure}

\begin{figure}
	\centering
	\includegraphics[width = 1.0\textwidth]{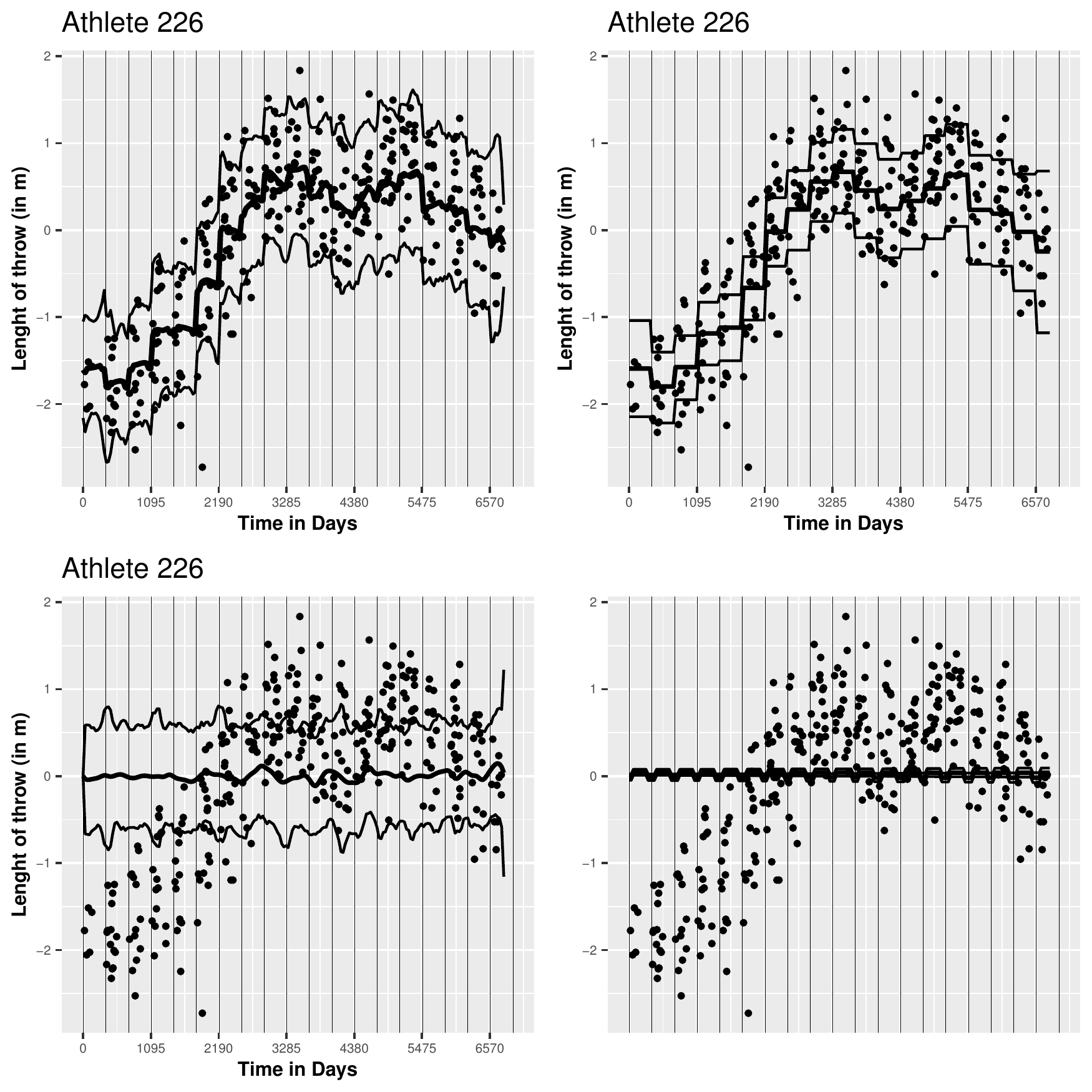}
	\caption{Single contributions to the whole additive model as in Equation \ref{compll}. The first panel is the complete additive model, whereas the second (top-right) displays the estimate of the seasonal random intercept. The third panel (bottom-left) represents the functional contribution, while the bottom-right panel displays the regressive component.}
	\label{comp}
\end{figure}

Performances for the different models are fairly similar: as a matter of fact, the model specifications do not differ in a significant way. Despite having a slightly lower LPML than the best performing model, $M_2$, we prefer looking at results for model $M_1$ with 80 degrees od freedom splines, GARCH errors and three regressors with time-dependent \textit{age} definition because regression parameters prove to be significant in this setting. As far as the estimation of trajectories describing the evolution of athletes' performances is concerned, we use the method discussed in Section \ref{estimation}. Figure \ref{Athl} displays the estimate (with $95 \%$ credible bounds) for a random selection of athletes (black) together with one-season-ahead performance prediction (grey). The results are graphically pleasing in terms of model fit, but some comments are of order. First, we acknowledge that the seasonal random intercept captures the majority of the variability in the data. Second, the functional component, which is meant to capture the overall variability in the data set, reduces to capture the intra-seasonal variability. Interestingly, the number on non-local bases selected by the adaptive procedure in \cite{shrink} is exactly equal to the number of seasons in the data set. This effect seems to be consistent with the choice of degrees of freedom, that limits the support of each spline to a unique season. Finally, the effect of covariates is very small in magnitude. Note that, because \textit{sex} is time-constant, in the estimated trajectory we expect to recognise the contribution of \textit{age} as a linear trend and the environmental effect as a diversification of summer and winter performances. \\
\indent In Figure \ref{comp} we underline the effect of the three additive components of our functional model. The first panel (top-left) displays the observed data and the estimated trajectory for athlete 226. The top-right panel shows the seasonal contribution (i.e, an estimation of $\mu_{is}; \ i = 226; \ s = 1,\ldots,S_{226}$). The third panel (bottom-left) reports the functional contribution (i.e, an estimate of $f_{226}(t)$). Finally, the bottom-right panel shows the effect of covariates. As anticipated, within the trajectory estimation of a unique athlete, we only recognise a performance drop predicted during winter months (negative effect of indoor environment on performances) in the bottom-right panel. The tight credible intervals for this component assures estimates to be significant, despite small, as we will discuss in further detail in the next Section.\\

\subsection{Parameters interpretation}

The regression parameters can be easily interpreted from a sports' analytics perspective. It is important to stress that, to improve convergence of the MCMC algorithm, we fitted our model centering athlete-specific data around their average. Accordingly, in our experiments the raw data $y_{ij}$ were substituted by the centered points:
\[ \tilde{y}_{ij} = y_{ij} - \frac{\sum_{j=1}^{n_i} y_{ij}}{n_i} = y_{ij} - \overline{y}_i \quad \text{for } i=1,\ldots,n  \text{ and } \ j=1,\ldots,n_i\]
as data-input for the model. We have to take into account this transformation when interpreting the regression
parameters, especially when dealing with dummy variables.\\
\indent We report the posterior mean estimate of the regression coefficients, their standard deviation, the effective sample size (ESS) and the $95 \%$ posterior credible bounds for the most interesting models. Table \ref{resM1} displays the results under model $M_1$, where covariates are \textit{sex} ($x_1$), \textit{age} ($x_2$) and \textit{environment} ($x_3$). 
 Even if the covariate effect is small in size, we observe that the $95\%$ credible intervals do not contain zero,
showing a significant effect. In particular $\beta_2$ is positive, suggesting that athletes, on average, increase their performances throughout their career. Further, $\beta_3$ is positive, meaning that an athlete is likely to perform better outdoors than indoors. Finally, $\beta_1$ is negative. Since \textit{sex} is a time-constant dummy variable, its coefficient quantifies the difference in variability of the athlete's performance around his/her average $\overline{y}_i$. We conclude that
female's trajectories express less variability around their average than men's. Parameter estimation under the other models considered in the paper are similar in sign with respect to the ones just discussed here. We report complete results in Appendix \ref{tables}. \\
\indent A final comment on results obtained using doping as additional regressor is required. We stressed that LPML performances for models $M_5$ and $M_6$ are quite low, however, this may be due to the fact that the data set is imbalanced, i.e., there are too few doped athletes (18 out of 653). In fact, the ESS of the parameter corresponding to \textit{doping} is very low. Nevertheless, it is interesting to observe that estimates are similar for all common parameters and that the coefficient of the doping regressor is
negative (even if the credible intervals contain zero). We conclude that the use of performance-enhancing drugs seems to have a negative effect on the variability of athletes' performances.

\begin{table}
	\centering
	\caption{Posterior mean estimate of the regression coefficients for model $M_1$(Table \ref{modelsummary}), together with the standard deviation of their estimate, effective sample size (ESS) with respect to 1600 retained samples, and $95 \%$ posterior credible bounds.}
	\begin{tabular}{lllllll} 
		\toprule
		Coeff & Mean & Sd & ESS & $2.5 \% $ & $97.5 \% $ \\ [0.5ex] 
		\midrule
		$\beta_1$ & -0.120 & 0.0270 & 190 & -0.175 & -0.0675 \\ 
		$\beta_2$ & 6.22 e-03 & 9.95 e-04 & 170 & 4.20 e-03 & 8.20 e-03 \\ 
		$\beta_3$ & 0.0453 & 9.55 e-03  & 1600 & 0.0269 &  0.0643 \\
		\bottomrule
	\end{tabular}
	\label{resM1}
\end{table}

\section{Discussion}

We proposed an additive hierarchical Bayesian model for the analysis of athletes' performances in a longitudinal context. Following \cite{MH}, we proposed a smooth functional contribution for explaining the overall variability in the data set. The functions are represented by means of a high-dimensional set of pre-specified basis functions and a factor model on the basis coefficients ensures dimensionality reduction. We enriched the model by allowing for time-dependent covariates to affect estimates through a regressive component. Finally, we addressed the issue of seasonal gathering of sports data introducing a mixed effect model with GARCH errors which provides evolving random intercepts over different time intervals in the data set. While the motivation of our work comes from the analysis of shot put performance data, the methodology presented in this work is applicable to the analysis of performance data collected in all measurable sports. \\
\indent The Bayesian latent factor methodology was originally developed for very sparse longitudinal
data, with the purpose of capturing a global trend in subject-specific trajectories. We balanced the model with the requirement of smoothness using a B-spline basis system and adding a seasonal random intercept. However, it is evident that the latter explains the majority of variability in the dataset. Therefore, it might be worth considering a functional basis that batter captures the intra-seasonal variability. Further, we observed that the contribution of the regressive component is consistent across various modelling choices. \\
\indent Finally, we looked at the effect of \textit{doping} on results. Despite not being significant, the negative effect seems to suggest that performances of doped athletes are less variable. This is in line with previous literature suggesting that doping is more likely used to enhance performances in periods of decreasing fitness than to consolidate already good performances, generally exposed to strict controls. We think this aspect deserves further investigation, considering, for instance, more specific modeling techniques and a less imbalanced data set.\\
\indent In conclusion, the attempt to extend exiting tools of functional statistics to the modelling of (shot put) performance data seems promising because of the adaptability of these methodologies to all sorts of performance longitudinal data in measurable sports.

\section*{Acknowledgements}
We would like to thank Prof. James Hopker (University of Kent, School of Sport and Exercise Sciences) for sharing the shot put data set, and for the helpful comments.

%
\section*{Conflict of interest}
The authors declare that the research was conducted in the absence of any commercial or financial relationships that could be construed as a potential conflict of interest.

\section*{Availability of data and material}
The raw data set is available at \url{www.tilastopaja.eu}, whereas the data prepared for our analysis at \url{https://github.com/PatricDolmeta/Bayesian-GARCH-Modeling-of-Functional-Sports-Data.}   

\section*{Code availability}
The code for the Bayesian analysis and the output production (numerical and graphical) is available at \url{https://github.com/PatricDolmeta/Bayesian-GARCH-Modeling-of-Functional-Sports-Data.}   



\bibliographystyle{apa}
\bibliography{literature}

\begin{appendices}
	\section{Posterior computations}
	\label{appendix}
	
	In this Appendix, we discuss in greater detail the full conditional updates anticipated in Section \ref{Update} and referred to in Algorithm \ref{Gibbs}. 
	
	\subsection{Functional model}
	With reference to the hierarchical model discussed in Section \ref{funpart}, we get the following updates:
	\begin{enumerate}
		\item Recalling that $\boldsymbol{\Lambda}$ is a $q \times k$ matrix, we update its $j$-th row as:
		\begin{gather*}  \pi(\boldsymbol{\lambda_j}|\ldots) \sim N_k((\boldsymbol{D^{-1}_j} + \sigma_j^{-2} \boldsymbol{\eta^\top} \boldsymbol{\eta}) ^{-1} \boldsymbol{\eta^\top} \sigma_j^{-2} \boldsymbol{\theta}^{(j)}, (\boldsymbol{D^{-1}_j} + \sigma_j^{-2} \boldsymbol{\eta^\top} \boldsymbol{\eta}) ^{-1} ) \\
			\text{where} \\
			\boldsymbol{D^{-1}_j} = diag(\phi_{j1}^{-1} \tau_1^{-1}, \ldots, \phi_{jk}^{-1} \tau_k^{-1})\\
			\boldsymbol{\eta^\top} = ( \eta_1, \ldots, \eta_k )\\
			\boldsymbol{\theta}^{(j)} = (\theta_{j1}, \ldots, \theta_{jn})
		\end{gather*}
		\item Concerning the local shrinkage parameters, we get:
		\[ \pi(\phi_{jh}|\ldots) \sim Ga \biggl(\frac{\nu_{\phi} + 1}{2},\frac{\nu_{\phi} + \tau_h \lambda_{jh}^2}{2} \biggr) \]
		\item The first global shrinkage parameter is updated as follows:\[ \pi(\delta_{1}|\ldots) \sim Ga \biggl(a_1 + \frac{q k}{2}, 1+ \frac{1}{2} \sum_{l=1}^{k} \tau_l^{(1)} \sum_{j=1}^{q} \phi_{jl}\lambda_{jl}^2 \biggr) \]
		\item Whereas for the remaining ones:
		\begin{gather*}
			\pi(\delta_{h}|\ldots) \sim Ga \biggl(a_h + \frac{q (k-h-1)}{2}, 1+ \frac{1}{2} \sum_{l=1}^{k} \tau_l^{(h)} \sum_{j=1}^{q} \phi_{jl}\lambda_{jl}^2 \biggr) \\
			\text{with} \\
			\tau_l^{(h)} = \prod_{t=1, t\neq h}^l \delta_t
		\end{gather*}
		\item Variances of factor residuals can be sampled as:
		\[ \pi(\sigma_{j}^{-2}|\ldots) \sim Ga \biggl(a_{\sigma} + \frac{n}{2}, b_{\sigma} + \frac{ \sum_{i=1}^{n} (\boldsymbol{\theta}^{(j)} - \boldsymbol{\Lambda} \eta_i)^2 }{2}\biggr) \]
		\item Latent factors for a specific individual have conditional posterior distribution: \begin{gather*}  \pi(\boldsymbol{\eta}_i|\ldots) \sim N (\boldsymbol{A_i}^{-1} \times \boldsymbol{B_i},\boldsymbol{A_i}^{-1} ) \\ 
			\text{being} \\
			\boldsymbol{A_i} = \boldsymbol{\Lambda}^\top \boldsymbol{B_i}^\top (\psi^2 \boldsymbol{I}_{n_i} + \boldsymbol{B_i} \boldsymbol{\Sigma} \boldsymbol{B_i}^\top)^{-1} \boldsymbol{B_i} \boldsymbol{\Lambda} + I_k \\
			\boldsymbol{B_i} = \boldsymbol{\boldsymbol{\Lambda}}^\top \boldsymbol{B_i}^\top (\psi^2 \boldsymbol{I}_{n_i} + \boldsymbol{B_i} \boldsymbol{\Sigma} \boldsymbol{B_i}^\top)^{-1} y_i
		\end{gather*}
		\item Finally, coefficients of the spline representation are sampled from:\begin{gather*}  \pi(\boldsymbol{\theta}_i|\ldots) \sim N (\boldsymbol{C_i}^{-1} \times \boldsymbol{D_i},\boldsymbol{C_i}^{-1} ) \\
			\text{being} \\
			\boldsymbol{C_i} = \psi^2 \boldsymbol{B_i}^\top \boldsymbol{B_i} + \boldsymbol{\Sigma}^{-1} \\
			\boldsymbol{D_i} = \psi^2 \boldsymbol{B_i}^\top y_i + \boldsymbol{\Sigma}^{-1} \boldsymbol{\Lambda} \boldsymbol{\eta}_{i}
		\end{gather*}
	\end{enumerate}

	\subsection{Seasonal GARCH model}
	Concerning the seasonal component, let us start noticing that the likelihood function of the GARCH model is:
	\begin{equation}
		f ( \  \{ \mu_{i,s} \}_{i=1,\ldots,n}^{s=1,\ldots,S_i} \ | \ \varpi, h_{i,s}) = \prod_{i=1}^n \prod_{s=1}^{S_i} \frac{1}{\sqrt{2 \pi h_{i,s}}} exp \left\{ - \frac{ \stackrel{\zeta_{i,s}^2}{\overbrace{(\mu_{i,s} - m)^2}}} {2 h_{i,s}} \right\}
	\end{equation}
	whereas the terms in the joint sampling distribution depending on $\mu_{i,s}$ are:
	\begin{gather*}
		\prod_{i=1}^n \prod_{s=1}^{S_i} \prod_{j \in s} \frac{1}{\sqrt{2 \pi \psi^2}} exp \left\{ - \frac{(y_{i,j}^{(2)} - \mu_{i,s})^2}{2 \psi^2} \right\} \\
		\text{where}\\
		y_{i,j}^{(2)} = y_{i,j} - f_i(t_{i,j}) - \boldsymbol{x}(t_{i,j}) \boldsymbol{\beta}  
	\end{gather*}
	is a convenient writing for partial residuals, especially from the implementation point of view.\\
	Consequently, $\mu_{i,s}$ can be updated taking samples from:
	\begin{align*}
		q_{\mu}(\mu_{i,s} \ | \ \varpi, y_{i,j}^{(2)}, \psi^2) & \propto exp \left\{ - \sum_{j \in s}  \frac{ \mu_{i,s}^2 - 2 \mu_{i,s} y_{i,j}^{(2)} } {2 \psi^2} \right\} \times exp \left\{ - \frac{\mu_{i,s}^2 - 2 \mu_{i,s} m } {2 h_{i,s}} \right\} \\
		& \sim N(\hat{\mu}_{\mu}, \hat{\Sigma}_{\mu})  \\
		\text{Where } \hat{\Sigma}_{\mu} = \biggl(  \sum_{j \in s} &  \frac{1}{\psi^2} + \frac{1}{h_{i,s}} \biggr)^{-1}  \text{ and } \  \hat{\mu}_{\mu} = \hat{\Sigma}_{\mu}^{-1} \biggl( \sum_{j \in s}  \frac{y_{i,j}^{(2)}}{\psi^2} + \frac{m}{h_{i,s}} \biggr)
	\end{align*}
	Posterior updates for other parameters are not straightforward in general because of the recursive definition of the conditional variance. As a consequence, we will rely on Metropolis Hastings steps within a Gibbs sampling algorithm. We iteratively generate:
	\begin{gather*}
		m^{[j]} \sim p(m \ | \ \alpha^{[j-1]}, \varpi^{[j-1]}, \mu_{i,s} ) \\
		\alpha^{[j]} \sim p(\alpha \ | \ m^{[j]}, \varpi^{[j-1]}, \mu_{i,s} ) \\
		\varpi^{[j]} \sim p(\varpi \ | \ m^{[j]},  \alpha^{[j]}, \mu_{i,s} )
	\end{gather*}
	
	\subsubsection{Update $m$}
	The posterior distribution for $m$ is based on the GARCH model under the assumption that the conditional variances $ \{ h_{i,s} \}_i^s $ are fixed, known and given by the writing in Eq. \eqref{garch}. In this case the likelihood function is:
	\begin{equation*}
		f ( \  \{ \mu_{i,s} \}_{i=1,\ldots,n}^{s=1,\ldots,S_i} \ | \ \boldsymbol{\alpha}, \varpi) = \prod_{i=1}^n \prod_{s=1}^{S_i} \frac{1}{\sqrt{2 \pi h_{i,s}}} exp \left\{ - \frac{ ( \mu_{i,s} - m)^2 } {2 h_{i,s}} \right\}
	\end{equation*}
	Accordingly, the posterior samples for $m$ will be drawn from:
	\begin{align*}
		q_m(m \ | \ \tilde{\boldsymbol{\alpha}}, \tilde{\varpi} ) & \propto exp \left\{ - \sum_{i=1}^n \sum_{s=1}^{S_i}  \frac{ m^2 - 2 m \mu_{i,s}  } {2 h_{i,s}} \right\} \times exp \left\{ - \frac{ m^2 - 2 m \mu_{m_0}  } {2 \Sigma_{m_0}} \right\} \\
		& \sim N(\hat{\mu}_m, \hat{\Sigma}_m)  \\
		\text{Where } \hat{\Sigma}_m = \biggl( & \sum_{i=1}^n  \sum_{s=1}^{S_i}  \frac{1}{h_{i,s}} + \frac{1}{\Sigma_{m_0}} \biggr)^{-1}  \text{ and } \  \hat{\mu}_m = \hat{\Sigma}_m^{-1} \biggl( \sum_{i=1}^n \sum_{s=1}^{S_i}  \frac{\mu_{i,s}}{h_{i,s}} + \frac{\mu_{m_0}}{\Sigma_{m_0}} \biggr)
	\end{align*}
	It is here of capital importance that $h_{i,s}$ are obtained by means of Eq. \eqref{garch}, using previous realisations the GARCH coefficients from the MH sampler.
	
	\subsubsection{Update $\boldsymbol{\alpha}$} 
	To generate samples from $\boldsymbol{\alpha}$ we can not exploit conjugacy, therefore we would like to rely on a normal proposal distribution. Indeed, symmetry of proposal distributions significantly eases computations of acceptance probabilities in Metropolis algorithms. Unfortunately, we are given a non-negativity constraint for the parameter, which forces us to apply a bidimensional transformation from the positive quadrant into the real plane: let say $(\theta_0, \theta_1)=(log(\alpha_0), log(\alpha_1))$. After transformation of the starting parameter, we propose a new sample employing a random walk sampler.\\
	\indent That is, a proposal of the form 
	\[ \boldsymbol{\theta}^{\ast} =\boldsymbol{\theta} + \boldsymbol{\zeta} \boldsymbol{\epsilon} \]
	where $\boldsymbol{\epsilon}$ is a multivariate standard normal random vector and $\boldsymbol{\zeta}$ a covariance matrix, defined according to an Adaptive scaling within the Adaptive Metropolis–Hastings algorithm (ASWAM) by \cite{HAARIO}. In this approach, both the covariance matrix of the proposal is adapted to the covariance matrix of the target density and the scale parameter is updated to achieve an average acceptance rate of 0.234 (proven to be optimal in different scenarios).\\
	\indent Provided it will be accepted, the actual parameter will be retrieved applying the inverse transform to $\boldsymbol{\theta}$. 
	As discussed above, the acceptance probability will not depend on the proposal distribution. Therefore:
	\begin{equation}
		\lambda_{\alpha} = min \left\{ \frac{p(\boldsymbol{\theta}^{\ast}\ | \ \varpi, \ m, \ \mu_{i,s})}{ p(\tilde{\boldsymbol{\theta}} \ | \ \varpi, \  m, \  \mu_{i,s}) }, 1 \right\} 
	\end{equation}
	With $p(\boldsymbol{\theta} \ | \ \varpi, m, \mu_{i,s})$ the full conditional density of $\boldsymbol{\theta}$ given the data and the rest of the parameters, which is clearly obtained from $p(\boldsymbol{\alpha} \ | \ \varpi, m, \mu_{i,s})$ by change of variables.
	
	\subsubsection{Update $\varpi$}         
	
	Similar arguments apply for $\varpi$. Samples will be proposed according to a normal distribution centered around $\tilde{\gamma} = log(\tilde{\varpi})$:
	\begin{equation*}
		\gamma^{\ast} \sim N( \gamma | \tilde{\gamma}, \Sigma_{\varpi})  
	\end{equation*}
	Here, $\Sigma_{\varpi}$ is defined, for each iteration $g$ as:
	\[ \sqrt{\Sigma_{\varpi}}^{(g+1)} = \zeta^{(g+1)} = \rho \bigl(  \zeta^{(g)} * w^{(g)} \bigl( \lambda_{\theta}^{(g)} - \overline{\lambda} \bigr) \bigr ) \]
	to achieve an average optimal acceptance rate of $\overline{\lambda} = 0.234$ \cite{HAARIO}. \\
	\indent Actual samples will be retrieved applying the inverse transform and accepted according to:
	\begin{equation}
		\lambda_{\varpi} = min \left\{ \frac{p(\gamma^{\ast}\ | \ \boldsymbol{\alpha}, \ m, \ \mu_{i,s})}{ p(\tilde{\gamma} \ | \ \boldsymbol{\alpha}, \  m, \  \mu_{i,s}) }, 1 \right\} 
	\end{equation}
	where again $p(\gamma \ | \ \boldsymbol{\alpha}, m, \mu_{i,s})$ is the full conditional density of $\gamma$ given by change of variables from $p(\varpi \ | \ \boldsymbol{\alpha}, m, \mu_{i,s})$
	
	\subsection{Multiple regression model}
	
	On the base of the prior settings and the overall sampling scheme \ref{full} we get:
	\begin{enumerate}
		\item Regression coefficient, jointly, as: \begin{gather*}  \pi(\boldsymbol{\beta}|\ldots) \sim N (A^{-1} \times B,A^{-1} ) \\
			\text{being} \\
			A = \boldsymbol{x^\top} \boldsymbol{x} / \psi^2 + \Sigma_{\boldsymbol{\beta}_0} \\
			B = \Sigma_{\boldsymbol{\beta}_0} \boldsymbol{\beta}_0 + \boldsymbol{x^\top} \boldsymbol{y^{(3)}} / \psi^2 
		\end{gather*}
		
		\item The variance hyperparameter instead is updated as:\[ \pi(\sigma_{\beta}^{-2}|\ldots) \sim Ga \biggl( \frac{N + \nu_{\beta}} {2}, \frac{ \nu_{\beta} \sigma_{\beta}^2 + \sum_{i=1}^{n} \sum_{j=1}^{n_i} (y_{ij} - f_i(t_{ij}) - \mu_{is} - \boldsymbol{x} \boldsymbol{\beta} )^2}{2} \biggr)  \]
	\end{enumerate}
	
	\subsection{Error term}
Finally, the full conditional posterior distribution for the model variance is:
	\begin{gather*}
		\pi(\psi^{-2}|\ldots) \sim Ga \biggl( \frac{N + \nu_{\psi}} {2}, \frac{ \nu_{\psi} \sigma_{\beta}^2 + \sum_{i=1}^{n} \sum_{j=1}^{n_i} \epsilon_{i,j}^2}{2} \biggr)\\
		\text{being}\\
		\epsilon_{i,j} = y_{i,j} - f_i(t_{i,j}) - \mu_{i,s} - \boldsymbol{x} \boldsymbol{\beta} 
	\end{gather*}

\section{Regression coefficients}
\label{tables}

\begin{table}[h!]
	\caption{Posterior mean estimate of the regression coefficients for model $M_2$(Table \ref{modelsummary}), together with the standard deviation of their estimate, effective sample size (ESS) with respect to 1600 retained samples, and $95 \%$ posterior credible bounds.}
	\centering
	\begin{tabular}{llllll}
		\toprule
		Coeff & Mean & Sd &ESS & $2.5 \% $ & $97.5 \% $ \\ [0.5ex] 
		\midrule
		$\beta_1$ & -5.58 e-03 & 0.0263 & 223 & -0.0585 & 2.00 e-03  \\ 
		$\beta_2$ & 3.34 e-04 & 1.16 e-03 & 107 & -2.60 e-03 & 2.00 e-03 \\ 
		$\beta_3$ & 0.0570 & 9.09 e-03 & 1600 & 0.0395 & 0.0743\\
		\bottomrule
	\end{tabular}
	\label{resM2}
\end{table}

\begin{table}
	\centering
	\caption{Posterior mean estimate of the regression coefficients for model $M_5$(Table \ref{modelsummary}), together with the standard deviation of their estimate, effective sample size (ESS) with respect to 600 retained samples, and $95 \%$ posterior credible bounds.}
	\begin{tabular}{llllll} 
		\toprule
		Coeff & Mean & Sd &ESS & $2.5 \% $ & $97.5 \% $ \\ [0.5ex] 
		\midrule
		$\beta_1$ & -0.120 & 0.0285 & 77 & -0.178 & -0.0690  \\ 
		$\beta_2$ &  6.71 e-03 & 1.11 e-03 & 64 & 4.40 e-3 & 8.10 e-3 \\ 
		$\beta_3$ & 0.0373 & 0.0108 & 600 & 0.0168 & 0.0580 \\
		$\beta_4$ & -0.116 & 0.0761 & 70 & -0.267 & -0.0365  \\
		\bottomrule
	\end{tabular}
	\label{resM5}
\end{table}

\begin{table}
	\centering
	\caption{Posterior mean estimate of the regression coefficients for model $M_6$(Table \ref{modelsummary}), together with the standard deviation of their estimate, effective sample size (ESS), and $95 \%$ posterior credible bounds.}
	\begin{tabular}{llllll} 
		\toprule
		Coeff & Mean & Sd &ESS & $2.5 \% $ & $97.5 \% $ \\ [0.5ex] 
		\midrule
		$\beta_1$ & -0.0104 & 0.0250 & 74 & -0.0617 & 0.0362   \\ 
		$\beta_2$ &  -4.60 e-04 & 1.14 e-03 & 67 & 2.80 e-03   & 1.90 e-03   \\ 
		$\beta_3$ & 0.0578 & 9.37 e-03 & 600 & 0.0408  & 0.0756  \\
		$\beta_4$ & -0.0103 & 0.0638 & 91 & -0.134  & 0.115  \\
		\bottomrule
	\end{tabular}
	\label{resM6}
\end{table}

\end{appendices}

\end{document}